\documentstyle[preprint,aps]{revtex}

\nobreak \penalty 5000
\def\str{\penalty-10000\hfilneg\ }      
\def\nostr{\hfill\penalty-10000\ }     

\def\u4s{\Upsilon_{4S}}

\def\z0{$Z^0$}

\def\alp{\relax\ifmmode \alpha_s\else $\alpha_s$\fi$\;$}
\def\alpmz{\relax\ifmmode \alpha_s(M_Z)\else $\alpha_s(M_Z)$\fi$\;$}
\def\alpmzsq{\relax\ifmmode \alpha_s(M_Z^2)\else
    $\alpha_s(M_Z^2)$\fi$\;$}

\def\zbbar{Z^0 \rightarrow b \overline{b}}

\font\bigprint=cmr12 scaled \magstep1
\font\medprint=cmr12
\font\small=cmr10

\begin{document}

\draft
\preprint{}

\vsize=8.5in

\vbox to 8.499in{
\thispagestyle{empty}
\null\vspace{-.75in}
 \begin{flushright} {\small\baselineskip 12pt plus 1pt
SLAC--PUB--6687\\ December 1994\\ (T/E)\\ }
\end{flushright}
\medprint
\smallskip
\vfil
\begin{center} {\Large A TEST OF THE FLAVOR INDEPENDENCE \\[.1in]
 OF STRONG  INTERACTIONS$^{\bigprint \ast}$}\\
\vspace{0.2in} {\bigprint  The SLD  Collaboration}$^\diamond$

\vspace{0.1in} {\bigprint Stanford Linear Accelerator Center}\\
{\bigprint Stanford University,  Stanford, CA 94309}\\
 \vspace{0.4in}

{\bigprint Abstract}
\end{center}

{\smallskip\par\narrower \narrower We present a comparison of the
strong couplings of light ($u$,~$d$,~and~$s$),
$c$, and $b$ quarks determined from multijet rates in flavor-tagged
samples of hadronic $Z^0$ decays recorded with the SLC Large Detector
at the SLAC Linear Collider.   Flavor separation on the basis of lifetime
and decay multiplicity differences among hadrons containing light, $c$,
and $b$ quarks was made using  the SLD precision tracking system.  We
find:
   $\alpha_s^{uds}/{\alpha_s^{\rm all}} = 0.987 \pm 0.027({\rm stat}) \pm
                 0.022({\rm syst}) \pm 0.022({\rm theory})$,
   $\alpha_s^c/{\alpha_s^{\rm all}} = 1.012 \pm 0.104 \pm
                 0.102 \pm 0.096$, and
   $\alpha_s^b/{\alpha_s^{\rm all}} = 1.026 \pm 0.041 \pm
                 0.041\pm 0.030.$
\par\smallskip}
\vfil
\vspace{0.2in}
\begin{center} Submitted to {\em Physical Review Letters}
\end{center}
\vfil
\vspace{0.1in}
\noindent
\footnoterule\noindent
\vbox{\footnotesize\baselineskip 11pt plus 1pt minus 1pt
\noindent
   $^\ast$Work supported  in part by Department of Energy
  contracts:
    DE--FG02--91ER40676 (BU),
   DE--FG03--92ER40701 (CIT),
   DE--FG03--93ER40788 (CSU),
 DE--FG02--91ER40672 (Col\-o\-rado),
  DE--FG02--91ER40677 (Ill-\i\-nois),
    DE--AC03-76SF00098 (LBL),
    DE--FG02--92ER40715 (Massa\-chu\-setts),
  DE--AC02--76ER03069 (MIT),
   DE--FG06--85ER40224 (Ore\-gon),
    DE--AC03--76SF00515 (SLAC),
   DE--FG05--91ER40627 (Ten\-nes\-see),
   DE--FG03--91ER40618 (UCSB),
    DE--FG03--92ER40689 (UCSC),
  DE--AC02--76ER00881 (Wis\-con\-sin),
   DE--FG02--92ER40704 (Yale);
    by National Science Foundation grants:
   PHY--89--21320 (Colum\-bia),
    PHY--92--04239 (Cin\-cin\-nati),
  PHY--88--17930 (Rut\-gers),
   PHY--91--13428 (UCSC),
 PHY--88--19316 (Van\-der\-bilt),
 PHY--92--03212 (Wash\-ing\-ton);
  by the Istituto Nazionale di Fisica Nucleare of Italy
 (Bologna, Ferrara, Frascati, Pisa, Padova, Perugia);
 by the Japan-US Cooperative Research Project on High Energy Physics
  (Nagoya, Tohoku);
  and by the UK Science and Engineering Research Council
 (Brunel and RAL).\\  \quad \\
$^\diamond$The SLD Collaboration authors and their  institutions are
listed following the References. }
\vspace{0.1in}}
\eject

\setcounter{page}{2}

A fundamental assumption of the theory of strong interactions, Quantum
Chromodynamics (QCD), is that the strong coupling $\alpha_s$ is
independent of quark flavor.  This can be tested by measuring the strong
coupling in events of the type $e^+e^-\rightarrow q{\bar q}(g)$ for
specific  quark flavors $q$. Although an absolute determination of
$\alpha_s$ for each quark flavor would have large theoretical
 uncertainties\cite{Acomp}, it is possible to test the
flavor-independence of QCD precisely by measuring ratios  of couplings
in which most experimental errors and theoretical uncertainties are
expected to cancel.   Since it has recently been  suggested\cite{TomR}
that a flavor-dependent anomalous quark chromomagnetic  moment could
modify the probability for the radiation of gluons,
 comparison of the strong	 coupling for different quark flavors may also
provide information  on physics beyond	the  Standard Model.

Comparisons of $\alpha_s$ for $b$ or $c$ quarks with $\alpha_s$ for all
flavors made at PETRA\cite{TASSO} were limited in precision to
$\pm 0.41$ ($c$) and $\pm 0.57$ ($b$)                 due to small data
samples and limited heavy quark  tagging capability. LEP measurements
of $\alpha^b_s/\alpha_s^{udsc}$
have
reached precisions between
$\pm 0.06$ and $\pm 0.02$ \cite{LEPalphas} under the assumption that
$\alpha_s$ is independent of                                  flavor for all
the
non-$b$ quarks.  The OPAL Collaboration has measured
$\alpha^f_s/\alpha_s^{\rm all}$  for all five flavors $f$ with no
assumption  on the relative value of $\alpha_s$ for different
flavors\cite{OPAL} to precisions of $\pm 0.026$ for $b$ and $\pm 0.09$
to $\pm 0.20$ for the other flavors. The kinematic signatures used to tag
$c$ and light quarks suffer from low efficiency and strong biases,  due to
preferential tagging of events without hard gluon radiation.

The SLC Large Detector (SLD) \cite{SLD} at the SLAC Linear Collider
(SLC) is an ideal environment in which to test the flavor independence of
strong interactions.
tracking capability of the Central Drift Chamber (CDC)\cite{CDC} and the
precision CCD Vertex Detector  (VXD)\cite{VXD},  combined with the
stable, micron-sized beam interaction point (IP), allows us to select
$\zbbar (g)$ and $Z^0 \rightarrow q_l \bar{q_l}(g)$
 ($q_l = u,\, d,\, s$) events using their quark decay lifetime signatures
with high efficiency and purity and low bias.   Here we present the first
precise                                                measurements of
$\alpha^b_s/\alpha_s^{\rm all}$,
$\alpha^c_s/\alpha_s^{\rm all}$, and
$\alpha^{uds}_s/\alpha_s^{\rm all}$ using this technique, and making no
assumptions about the relative values           of $\alpha^{b}_s$,
$\alpha^{c}_s$ and $\alpha^{uds}_s$.

This analysis is based on the 1.8 $pb^{-1}$ of
$e^+e^-$ annihilation data collected during the 1993 run of the SLD at the
SLC at a mean center-of-mass energy of $\sqrt{s}=91.26$ GeV. The
trigger and selection criteria for hadronic $Z^0$ decays are described  in
Ref.~\cite{Acomp}. The efficiency for selecting a well-contained $Z^0
\rightarrow q{\bar q}(g)$ event was estimated to be above 96\%
independent of quark flavor. The estimated $0.10 \pm 0.05\%$
background was dominated by $Z^0 \rightarrow \tau^+\tau^-$ events. This
analysis used charged tracks measured in the CDC and in the
VXD\cite{Acomp}.

We used normalized impact parameters $d/\sigma_d$ as the basis for
quark flavor tags, where $d$ is the signed distance of closest approach
of a  charged track to the IP in the ($x$--$y$) plane transverse to the
beam axis, and $\sigma_d$ is the error on $d$. A resolution on $d$ of
10.8~$\mu$m has been measured using $Z^0 \rightarrow \mu^+\mu^-$
decays, and the spatial resolution on the average transverse IP position
has been measured to be 7~$\mu$m \cite{Rb}.  The distributions of $d$
and $d/\sigma_d$  are modeled well by the SLD simulation\cite{Rb}.
Tracks used for event flavor tagging were required to have: at  least one
VXD hit; at least 40~CDC hits, with the first hit at a radius less than
39~cm; a combined CDC+VXD fit quality $\sqrt{2\chi ^2} - \sqrt{2n_{\rm
d.o.f.}-1} < 8.0$;  momentum   greater than  0.5~GeV/$c$;
$\sigma_d < 250~\mu$m; and to miss the IP by  less than 0.3 cm in the
$x$--$y$ plane and by  less than 1.5~cm in $z$. Tracks from candidate
$K^0$ and $\Lambda$
  decays and $\gamma$-conversions were removed.

Figure~1 shows the distribution of $n_{\rm sig}$, the number of tagging
tracks per event with $d/\sigma_d \geq 3$.  The data are well described
by a  Monte Carlo  simulation of  hadronic $Z^0$ decays\cite{JETSET}
with parameter values tuned\cite{Nb}
 to hadronic $e^+e^-$ annihilation data, combined with a simulation of the
SLD. For the simulation,  the contributions of events  of different quark
flavors are shown separately.  The leftmost bin contains predominantly
events containing primary $u$, $d$, or $s$ quarks, while the rightmost
bins  contain a pure sample of events containing primary $b$ quarks.  The
event sample was divided accordingly into three parts:  those events
with $n_{\rm sig}=0$ were defined  to be the $uds$-tagged sample; those
with $1 \leq n_{\rm sig}\leq 3$ were the $c$-tagged sample; and those
with $ n_{\rm sig}\geq 4$ were the $b$-tagged sample. The efficiencies
$\varepsilon$ for selecting events (after cuts) of type $i$ ($i=uds,\, c,\,
b$)  with tag $i$,  and the fractions $\Pi$ of events of type $i$ in the
$i$-tagged sample,  were calculated from the Monte Carlo simulation to
be:
$(\varepsilon,\Pi)_{uds} = (77\%,\, 86\%)$;
$(\varepsilon,\,\Pi)_c = (59\%,\, 38\%)$; and $(\varepsilon,\,\Pi)_b  =
(46\%,\, 94\%)$.

\medskip
\vspace*{5.15in}
\hspace*{-8pt}
\special {illustration 7843A01doc scaled 1150}
\vspace*{1pt}

{\smallskip\narrower\baselineskip 16pt\noindent  \noindent      Figure
1. The measured distribution of the number of tracks per event with
$d/\sigma_d \geq 3.0$ (points).
 The histograms show the flavor composition estimated from a Monte
Carlo simulation (see text).
\medskip\par}

Jets were then reconstructed using iterative clustering  algorithms. We
used the  `E', `E0', `P', and `P0' variations of the {\sc Jade} algorithm,  as
well as the `Durham' (`D') and `Geneva' (`G')  algorithms\cite{Siggi}. We
divided events into two categories: those containing:  (1) two jets, and
(2) three or more jets. The fraction of the event sample in category 2
was defined as the 3-jet rate $R_3$.   This quantity is infrared- and
colinear-safe and has been calculated to ${\cal O}(\alpha_s^2)$ in
perturbative  QCD \cite{Siggi,KN}. For each algorithm, the jet resolution
parameter $y_c$ was chosen to be as small as possible subject to the
requirement that ${\cal O}(\alpha_s^2)$ QCD provides a good description
of
$R_3$ measured in our global sample of all  flavors \cite{Acomp,alphas}.
This choice
  maximizes $R_3$ while avoiding the `Sudakov region' at low $y_c$
where  multiple gluon emission  requires that large logarithmic terms of
$1/y_c$ be resummed in order to describe the data \cite{Acomp}.   The
resulting $y_c$ values are listed in Table~1.
\medskip

\setlength{\tabcolsep}{.25mm}

\vbox{
\setlength{\tabcolsep}{.25mm}
\centering
\begin{tabular}{ccccccc}
\multicolumn{7}{c}{{\bf Table 1.}~ Results
 for $R_3^j/R_3^{\rm all}$,
  derived from Eq.~1; see text.   Errors shown are statistical.} \\[.025in]
\hline  Algorithm
 &  $y_c$
 &  $R_3^{uds}/R_3^{all}$
 &  $R_3^{c}/R_3^{all}$
 &  $R_3^{b}/R_3^{all}$
 & $R_3^{c}/R_3^{u}\,$factor~
 & $R_3^{b}/R_3^{d}\,$factor \\[.015in]
\hline  E
 & ~0.080~~
 &  ~~0.941$\pm$0.042~~
 &	~~1.212$\pm$0.173~~
 &	~~0.980$\pm$0.063~~
 &  0.995
 &  0.958 \\ E0
 &  0.050~
 &  0.975$\pm$0.036
 &  1.113$\pm$0.145
 &	0.981$\pm$0.053
 &  0.994
 &  0.945 \\ P
 &  0.030~
 &  1.001$\pm$0.027
 &  0.985$\pm$0.109
 &	 1.007$\pm$0.041
 &  0.992
 &  0.929  \\  P0
 &  0.030~
 &   1.014$\pm$0.026
 &   0.899$\pm$0.102
 &	 1.037$\pm $0.039
 &  0.992
 &  0.929  \\
   D
 &  0.015~
 &   0.989$\pm$0.035
 &  1.096$\pm$0.145
 &	0.947$\pm$0.049
 &  0.991
 &  0.921 \\
 G
 &  0.030~
 &  1.032$\pm$0.020
 &   0.942$\pm$0.079
 &	 0.952$\pm$0.030
 &  0.989
 &  0.915 \\
\hline\\
\end{tabular}}

The $R_3^j$ for each of the $j$ quark types ($j=uds,\, c,\, b$) was
extracted from a maximum likelihood fit to
$n_2^i$ and $n_3^i$, the number of 2-jet and 3-jet events, respectively,
in the $i$-tagged sample:
\begin{eqnarray}   &  n_{2}^i\ =&\ \sum_{j=1}^3
\left(\varepsilon_{(2\rightarrow 2)}^{ij}
 (1-R_3^j) + \varepsilon_{(3\rightarrow 2)}^{ij} R_3^j\right) f^j N
\nonumber  \\ & n_{3}^i\ =&\ \sum_{j=1}^3 \
\left( \varepsilon_{(3\rightarrow 3)}^{ij}
 R_3^j +  \varepsilon_{(2\rightarrow 3)}^{ij} (1-R_3^j)\right) f^j N \ .
 \end{eqnarray}  Here $N$ is the total number of selected events
corrected  for the event selection efficiency, and
$f^j$ is the Standard Model fractional hadronic width for $Z^0$  decays to
quark type~$j$. The matrices
$\varepsilon_{(2\rightarrow 2)}^{ij}$ and
$\varepsilon_{(3\rightarrow 3)}^{ij}$  are the efficiencies for an event
of type $j$, with  2- or 3-jets at
 the parton level, to pass all  cuts and be tagged as a 2- or 3-jet event,
respectively, of type~$i$. Matrices $\varepsilon_{(2\rightarrow 3)}^{ij}$
and
$\varepsilon_{(3\rightarrow 2)}^{ij}$  are the efficiencies for an event
of type~$j$, with  2- or 3-jets at the parton level,   to pass all cuts and
be tagged as a 3- or 2-jet event, respectively, of type~$i$. This
formalism explicitly accounts for modifications of the parton-level
3-jet rate due to hadronization, detector effects,  and tagging bias.
These matrices were calculated from the Monte Carlo simulation.   The
efficiencies for correctly tagging a 2-jet event and a 3-jet event differ
by an average of 5.7\%, 8.3\%, and 30.3\% for the
$uds$, $c$, and $b$ tags, respectively.

\looseness=-2 Equations 1 were solved using 3-jet events defined by
each of the six algorithms. The ratios $R_3^j/R_3^{\rm all}$,  where
$R_3^{\rm all}$ is the 3-jet rate in the total event sample, are shown in
Table 1. Averaged over all six algorithms the correlation coefficients
from the fit are:
$uds$-$c:-0.76$, $uds$-$b:0.30$, $c$-$b:-0.55$. The statistical errors
were calculated using the full covariance matrix.

The 3-jet rate in heavy quark ($b$, $c$) events is expected to be reduced
relative to that in light quark events by the diminished phase-space for
gluon emission due to the quark masses. We evaluated the suppression
factors,
$R_3^c/R_3^u$ and $R_3^b/R_3^d$, for each jet algorithm and $y_c$ value
according to Ref. ~\cite{Maina},   assuming $b$ ($c$) quark  masses of
4.75 GeV/$c^2$ (1.50 GeV/$c^2$). These factors are listed in Table~1,
 and were used to correct the measured 3-jet rate ratios.

To ${\cal O}(\alpha_s^2)$ in perturbative QCD,
$R_3(y_c) = A(y_c)\alpha_s +
\bigl(B(y_c)+C(y_c)\bigr)\alpha_s^2$,
 where the ${\cal O}(\alpha_s^2)$ coefficient includes a term $B(y_c)$
from 3-parton states calculated at next-to-leading order, and a term
$C(y_c)$ from 4-parton states calculated at leading order. Hence,  the
ratio of the strong coupling of quark type $j$ to the mean coupling in the
sample of all flavors,
$\alpha_s^j/\alpha_s^{\rm all}$, can be determined from:
\begin{eqnarray}  &{{\textstyle  R_3^j(y_c)}\over {\textstyle R_3^{\rm
all}(y_c)} }\  =\   &{{A(y_c)\,\alpha_s^j\ +\ \bigl[B(y_c)+C(y_c)\bigr]\
(\alpha_s^j)^2}
\over{A(y_c)\,\alpha_s^{\rm all}\
 + \ \bigl[ B(y_c)+C(y_c)\bigr]\  (\alpha_s^{\rm all})^2}}\ ,
\end{eqnarray} where $A(y_c)$, $B(y_c)$, and $C(y_c)$  for the different
jet-finding algorithms were evaluated using  Refs.~\cite{Siggi,KN}.
Using our measured values of $\alpha_s^{\rm all}(M_Z^2)$
 determined from jet  rates \cite{alphas}, we found that for the E, E0, P,
P0, and D algorithms,  the  leading-order QCD calculation $C(y_c)
\alpha_s^2$ lies below the experimental 4-jet rate by roughly a factor of
two. We increased
$C(y_c)$   {\it ad hoc}\/\  for these algorithms, so as to describe
 the data. Equation~2 was solved to obtain $\alpha_s^j/\alpha_s^{\rm
all}$  for each jet algorithm;   the results are shown in Fig.~2.  The
errors include contributions from the statistical error, as well as the
experimental
 systematic errors and theoretical uncertainties.

We considered systematic effects that could modify the tagging
efficiencies. In each case   the error was evaluated by varying the
appropriate parameter in the Monte Carlo simulation,  recalculating the
matrices
$\varepsilon$, performing a new fit to Eq.~1, and rederiving
 $\alpha_s^j/\alpha_s^{\rm all}$. Suitable variation about the world
average value of each parameter was  considered\cite{Rb}. The
errors\str\eject

\medskip
\vspace*{4.8in}
\hspace*{.85in}
\special {illustration 7843A02doc scaled 950}
\vspace*{4pt}

{\smallskip\narrower\baselineskip 16pt\noindent     Figure 2.  Values of
$\alpha_s^j/\alpha_s^{\rm all}$ derived for each of the six jet
algorithms for each of the quark flavors~$j$ (see text). The error bars on
the averages include  the statistical and systematic errors and the total
theoretical uncertainty.
\medskip\par}

\noindent are summarized in Table~2, where averages over the six
algorithms are shown.   The largest
 contributions result from limited knowledge  of the heavy quark
fragmentation functions and
  $B$ decay multiplicity.  The uncertainty in $BR(Z^0 \rightarrow c \bar
c)$ also produces large variations in  $\alpha_s^c/\alpha_s^{\rm all}$
and $\alpha_s^{uds}/\alpha_s^{\rm all}$.  Contributions from $b$ hadron
lifetimes, the fraction of $D^+$ in $B$ meson decays, $b$ baryon
production rates, and the charm hadron decay multiplicity are small. The
detector systematic error is dominated by  the uncertainty in the
charged track reconstruction efficiency. No systematic variation of the
results was found when the event selection cuts, tag criteria, or $y_c$
values were changed.
\eject

\vbox{\centering
\begin{tabular}{lccc}
\multicolumn{4}{c}{{\bf Table~2. }\  Contributions to the systematic
error on
$\textstyle \alpha_s^j\,/ \,\textstyle \alpha_s^{\rm all}$.} \\[.12in]
\hline\\[-.25in]
 Source
 &  $~~~\Delta \left( {\textstyle \alpha_s^{uds}\,/\,
 {\textstyle \alpha_s^{\rm all}}} \right)~~~$
 &  $~~~\Delta \left( {\textstyle \alpha_s^c\,/\, {\textstyle
\alpha_s^{\rm all}}} \right)~~~$
 &  $~~~\Delta \left( {\textstyle \alpha_s^b\,/\, {\textstyle
\alpha_s^{\rm all}}} \right)~~~$ \\[.12in]
\hline
$b$ physics           &  0.008  &	0.060	&  0.033         \\
$c$ physics           &  0.017  &	0.060	&  0.011         \\ Detector
modeling   &  0.003  &	0.032	&  0.017         \\ Monte Carlo
statistics~~~         &  0.011   &	0.048	&  0.014         \\ QCD
uncertainty     &  0.003  &	0.011	&  0.012 \\
\hline\\
\end{tabular}}

We considered sources of uncertainty in the QCD  predictions that affect
the values of $\alpha_s^j/\alpha_s^{\rm all}$ derived from Eq.~2.  For
each jet algorithm these  include variation of the QCD renormalization
scale within the range allowed by our measurements  of jet rates in the
global sample \cite{alphas} and variation of the heavy quark masses used
in the phase-space correction factors by $\pm 0.25$~GeV/$c^2$.  In
addition, the shifts in
$\alpha_s^j/\alpha_s^{\rm all}$ due to the {\it ad hoc} increase of the
coefficient $C(y_c)$ were conservatively assigned as an uncertainty. The
variation of the results due to uncertainties in parton production and
hadronization were found to be small.  These contributions were added in
quadrature to yield the total QCD uncertainties listed in
Table~2.

There is significant scatter among the
$\alpha_s^j/\alpha_s^{\rm all}$ values derived from the different jet
algorithms.  In order to quote a single $\alpha_s^j/\alpha_s^{\rm all}$
value for each flavor $j$,  we made the conservative assumption that the
results are completely correlated,
 and we calculated the unweighted mean values and errors over all six
algorithms. We obtained
\begin{eqnarray}  &{{ \textstyle \alpha_s^{uds}}\over {\textstyle
\alpha_s^{\rm all}}}\  =&\ 0.987 \pm 0.027\,({\rm stat}) \pm
                 0.022\,({\rm syst}) \pm 0.022\,({\rm theory})\ ,
\nonumber \\[.1in]
  & {{\textstyle \alpha_s^c }\over {\textstyle \alpha_s^{\rm all}}}\  =&\
1.012 \pm 0.104\,({\rm stat}) \pm
                 0.102\,({\rm syst}) \pm 0.096\,({\rm theory})\ ,
\nonumber \\[.1in]
 &  {{\textstyle \alpha_s^b }\over{\textstyle \alpha_s^{\rm all}}}\  =&\
1.026 \pm 0.041\,({\rm stat}) \pm
                 0.041\,({\rm syst})\pm 0.030\,({\rm theory})\ ,
\end{eqnarray}  where the theoretical uncertainty is the sum in
quadrature  of the QCD uncertainty from Table~2  and  the  rms of the
results  over the six algorithms. These averages are also shown in
Fig.~2.   The variation of results among jet algorithms, presumably due
to               different uncalculated ${\cal O}(\alpha_s^3)$ QCD
contributions,                           dominates the theoretical uncertainty,
is
not small compared  with experimental errors, and has             not been
considered in previous analyses \cite{LEPalphas,OPAL}.

In conclusion, we have used hadron lifetime information  to separate
hadronic $Z^0$ decays into three flavor samples with high efficiency and
purity,  and small
 bias against events containing hard gluon radiation. From a comparison
of the rates of multijet events in these samples, we find that the strong
coupling is independent of quark flavor within our sensitivity.  These are
the first such results using a precision vertex detector  for flavor
separation at the~$Z^0$. This represents  the most precise test for $uds$
events. Our findings are consistent with measurements performed at LEP
using different flavor-tagging  techniques \cite{LEPalphas,OPAL}.

We thank the personnel of the SLAC accelerator department and the
technical staffs of our collaborating institutions for their outstanding
efforts.
 We thank C.~Ng, T.~Rizzo, and E.~Maina
 for their helpful contributions.

\eject
 \def\iADEL{$^{(1)}$}
  \def\iBOL{$^{(2)}$}
  \def\iBU{$^{(3)}$}
  \def\iBRUN{$^{(4)}$}
  \def\iCIT{$^{(5)}$}
  \def\iUCSB{$^{(6)}$}
  \def\iUCSC{$^{(7)}$}
  \def\iCIN{$^{(8)}$}
  \def\iCSU{$^{(9)}$}
  \def\iCOLO{$^{(10)}$}
  \def\iCOL{$^{(11)}$}
  \def\iFER{$^{(12)}$}
  \def\iFRA{$^{(13)}$}
  \def\iILL{$^{(14)}$}
  \def\iLBL{$^{(15)}$}
  \def\iMIT{$^{(16)}$}
  \def\iMASS{$^{(17)}$}
  \def\iMISS{$^{(18)}$}
  \def\iNAG{$^{(19)}$}
  \def\iOREG{$^{(20)}$}
  \def\iPAD{$^{(21)}$}
  \def\iPERU{$^{(22)}$}
  \def\iPISA{$^{(23)}$}
  \def\iRUT{$^{(24)}$}
  \def\iRAL{$^{(25)}$}
  \def\iSOGANG{$^{(26)}$}
  \def\iSLAC{$^{(27)}$}
  \def\iTENN{$^{(28)}$}
  \def\iTOH{$^{(29)}$}
  \def\iVAND{$^{(30)}$}
  \def\iWASH{$^{(31)}$}
  \def\iWISC{$^{(32)}$}
  \def\iYALE{$^{(33)}$}
  \def\dead{$^{\dag}$}
  \def\andgen{$^{(a)}$}
  \def\andper{$^{(b)}$}

\begin{center}
%
%
%

%
%

 \section*{THE SLD COLLABORATION}
\medprint\baselineskip 15pt plus 1pt minus1pt

\mbox{K. Abe                 \unskip,\iTOH}
\mbox{I. Abt                 \unskip,\iILL}
\mbox{C.J. Ahn               \unskip,\iSOGANG}
\mbox{T. Akagi               \unskip,\iSLAC}
\mbox{N.J. Allen             \unskip,\iSLAC}
\mbox{W.W. Ash               \unskip,\iSLAC$^\dagger$}
\mbox{D. Aston               \unskip,\iSLAC}
\mbox{N. Bacchetta           \unskip,\iPAD}
\mbox{K.G. Baird             \unskip,\iRUT}
\mbox{C. Baltay              \unskip,\iYALE
% ...
}
\eject
\end{center}

\baselineskip = 18pt plus 1pt minus 1pt
\leftskip .25in\it
 \def\iADEL{$^{(1)}$}
  \def\iBOL{$^{(2)}$}
  \def\iBU{$^{(3)}$}
  \def\iBRUN{$^{(4)}$}
  \def\iCIT{$^{(5)}$}
  \def\iUCSB{$^{(6)}$}
  \def\iUCSC{$^{(7)}$}
  \def\iCIN{$^{(8)}$}
  \def\iCSU{$^{(9)}$}
  \def\iCOLO{$^{(10)}$}
  \def\iCOL{$^{(11)}$}
  \def\iFER{$^{(12)}$}
  \def\iFRA{$^{(13)}$}
  \def\iILL{$^{(14)}$}
  \def\iLBL{$^{(15)}$}
  \def\iMIT{$^{(16)}$}
  \def\iMASS{$^{(17)}$}
  \def\iMISS{$^{(18)}$}
  \def\iNAG{$^{(19)}$}
  \def\iOREG{$^{(20)}$}
  \def\iPAD{$^{(21)}$}
  \def\iPERU{$^{(22)}$}
  \def\iPISA{$^{(23)}$}
  \def\iRUT{$^{(24)}$}
  \def\iRAL{$^{(25)}$}
  \def\iSOGANG{$^{(26)}$}
  \def\iSLAC{$^{(27)}$}
  \def\iTENN{$^{(28)}$}
  \def\iTOH{$^{(29)}$}
  \def\iVAND{$^{(30)}$}
  \def\iWASH{$^{(31)}$}
  \def\iWISC{$^{(32)}$}
  \def\iYALE{$^{(33)}$}
  \def\dead{$^{\dag}$}
  \def\andgen{$^{(a)}$}
  \def\andper{$^{(b)}$}

\noindent
  \iADEL
     Adelphi University,
     Garden City, New York 11530 \nostr
 \iBOL
     INFN Sezione di Bologna,
     I-40126 Bologna, Italy \nostr
  \iBU
     Boston University,
     Boston, Massachusetts 02215\nostr
  \iBRUN
     Brunel University,
     Uxbridge, Middlesex UB8 3PH, United Kingdom \nostr
  \iCIT
     California Institute of Technology,
     Pasadena, California 91125 \nostr
  \iUCSB
     University of California at Santa Barbara,
     Santa Barbara, California 93106 \nostr
  \iUCSC
     University of California at Santa Cruz,
     Santa Cruz, California 95064 \nostr
  \iCIN
     University of Cincinnati,
     Cincinnati, Ohio 45221\nostr
  \iCSU
     Colorado State University,
     Fort Collins, Colorado 80523 \nostr
  \iCOLO
     University of Colorado,
     Boulder, Colorado 80309 \nostr
  \iCOL
     Columbia University,
     New York, New York 10027 \nostr
  \iFER
     INFN Sezione di Ferrara and Universit\`a di Ferrara,
     I-44100 Ferrara, Italy \nostr
  \iFRA
     INFN  Lab. Nazionali di Frascati,
     I-00044 Frascati, Italy \nostr
  \iILL
     University of Illinois,
     Urbana, Illinois 61801 \nostr
  \iLBL
     Lawrence Berkeley Laboratory, University of California,
     Berkeley, California 94720 \nostr
  \iMIT
     Massachusetts Institute of Technology,
     Cambridge, Massachusetts 02139 \nostr
  \iMASS
     University of Massachusetts,
     Amherst, Massachusetts 01003 \nostr
  \iMISS
     University of Mississippi,
     University, Mississippi  38677 \nostr
  \iNAG
     Nagoya University,
     Chikusa-ku, Nagoya 464 Japan  \nostr
  \iOREG
     University of Oregon,
     Eugene, Oregon 97403 \nostr
  \iPAD
     INFN Sezione di Padova and Universit\`a di Padova,
     I-35100 Padova, Italy\nostr
  \iPERU
     INFN Sezione di Perugia and Universit\`a di Perugia,
     I-06100 Perugia, Italy \nostr
  \iPISA
     INFN Sezione di Pisa and Universit\`a di Pisa,
     I-56100 Pisa, Italy \nostr
  \iRUT
     Rutgers University,
     Piscataway, New Jersey 08855 \nostr
  \iRAL
     Rutherford Appleton Laboratory,
     Chilton, Didcot, Oxon OX11 0QX United Kingdom \nostr
  \iSOGANG
     Sogang University,
     Seoul, Korea \nostr
  \iSLAC
     Stanford Linear Accelerator Center, Stanford University,
     Stanford, California 94309 \nostr
  \iTENN
     University of Tennessee,
     Knoxville, Tennessee 37996 \nostr
  \iTOH
     Tohoku University,
     Sendai 980 Japan \nostr
  \iVAND
     Vanderbilt University,
     Nashville, Tennessee 37235 \nostr
  \iWASH
     University of Washington,
     Seattle, Washington 98195 \nostr
  \iWISC
     University of Wisconsin,
     Madison, Wisconsin 53706 \nostr
  \iYALE
     Yale University,
     New Haven, Connecticut 06511 \nostr
  \dead
     Deceased \break
  \andgen
     Also at the Universit\`a di Genova \nostr
  \andper
     Also at the Universit\`a di Perugia \nostr
\rm

\end{document}